\documentclass[preprint,showpacs,preprintnumbers,amsmath,amssymb]{revtex4}
\usepackage{graphicx}
\usepackage{dcolumn}
\usepackage{bm}
\usepackage{epsfig}
\usepackage{subfigure}

\newcommand{\bi}{\begin{itemize}}
\newcommand{\ei}{\end{itemize}}
\newcommand{\be}{\begin{eqnarray}}
\newcommand{\ee}{\end{eqnarray}}
\newcommand{\bbmatrix}{\left( \begin{array}}
\newcommand{\eematrix}{\end{array} \right)}

\begin{document}

\title{LaMnO$_3$ is a Mott Insulator: a precise definition and an evaluation of the local interaction strength}
\author{Chungwei Lin and Andrew.~J.~Millis}
\affiliation{ Department of Physics, Columbia University \\
538W 120th St NY, NY 10027}

\begin{abstract}
We compare the interaction parameters measured on LaMnO$_3$ to single site dynamical mean field estimates of the
critical correlation strength needed to drive a Mott transition, finding that the total correlation strength
(electron-electron plus electron-lattice) is very close to but slightly larger than the critical value, while
if the electron lattice interaction is neglected the model is metallic. Our results emphasize the importance
of additional physics including the buckling of the Mn-O-Mn bonds.
\end{abstract}

\maketitle



The ``colossal'' magnetoresistance (CMR) materials are widely regarded as paradigm ``strongly correlated''
systems in which strong local interactions combine with orbital and lattice effects to
produce a range of exotic behavior \cite{Imada_98}. Many papers over the last  
decade have argued that LaMnO$_3$, the ``parent compound'' of the CMR family, is a strongly correlated Mott  
insulator \cite{Imada_98, Kovaleva_04, Held_00, Yang_07}, while others have argued that short or
long ranged Jahn-Teller order is the key physics \cite{Millis_96-2, AhnMillis_00}. The debate has continued in the
literature up to the present. 

Two classes of difficulty have complicated the discussion of LaMnO$_3$. The first concerns the definition
of ``Mott insulator''. 
While there is a general agreement that the term ``Mott insulator'' \cite{Mott_37,Imada_98} refers to materials in which the
electronic correlations are strong enough to lead to insulating behaviors in the absence of long ranged
order or significant short ranged order, it is not easy to implement this definition in practice since most candidate Mott insulators 
exhibit some form of long ranged order at low temperature (in LaMnO$_3$ ``low temperature'' means below the 
orbital ordering temperature $T_{oo} \sim 750K$) 
and neither long ranged order nor any short ranged correlation can easily be ``turned off'' experimentally.
The second class of difficulty concerns the ambiguity of the concept of ``correlation strength''. The
paradigmatic model for Mott insulator is the one orbital Hubbard model in which there is just one interaction parameter.
The rich multiplet structure associated with partially filled d-shells in materials such as LaMnO$_3$ includes intra and
inter orbital Coulomb repulsions, Hund's couplings and Jahn-Teller splittings. Understanding which interaction is the
most important has not been easy. 
These difficulties can be circumvented theoretically. In a theory the effects of different interactions
can be disentangled. Further the single site
dynamical mean field theory (S-DMFT) \cite{DMFT_96} neglects intersite  
correlations entirely, but produces a metal-insulator transition for interactions larger than a critical  
value. We propose that a material is  ``strongly correlated'' if, for the relevant  
interaction parameters, the single site DMFT approximation produces an insulating solution at zero temperature 
without long-ranged order.
Materials with weaker but still non-negligible correlations should be  
referred to as having intermediate correlations.

In this paper we combine experimental and theoretical information to show that the local interactions in LaMnO$_3$ are 
very close to the critical values needed to drive a metal-insulator transition in the single site DMFT approximation and in the
absence of long ranged or short-ranged order. 
While aspects of our analysis have appeared in the literature, in this paper we present a comprehensive view
which allows us to resolve a dispute which has continued in the literature up to the present.
The close proximity of the interactions to the critical value means that the behavior is extremely
sensitive to the conduction bandwidth, to magnetic order, and to the electron-lattice coupling.
In particular, for the bandwidth obtained from local density approximations for the observed structure the material is
insulating in the paramagnetic phase but would be metallic in a hypothetically ferromagnetic
phase or if the electron-phonon coupling were set to zero. Further, the observed crystal structure 
(at high temperatures above the orbital ordering temperature) differs from the ideal
perovskite structure by a GdFeO$_3$ rotation. If this rotation is removed, the material becomes metallic.

The rest of the paper is organized as follows. We first establish the Hamiltonian and present qualitative considerations.
We next present an analysis of the changes in optical conductivity across the Neel transition which, 
following Ref\cite{Kovaleva_04} and Ref\cite{AhnMillis_00}, allows us to estimate the interaction strength.
We then present single site DMFT calculations for various model parameters which allow us to estimate
the proximity of the parameters to the threshold values. We conclude with a summary and a general classification scheme. 

We now estimate the electronic parameters, beginning with the bandwidth $W$. 
LDA and LDA+U calculations reveal that the $e_g$ bands are well described by a nearest 
neighbor tight binding model with hopping amplitude $t$ \cite{Ederer_07}; the $e_g$ symmetry implies a directional
structure to the hopping so that there are two bands and $W=6t$.
The bandwidth is found to depend strongly on 
the crystal structure; in particular on the magnitude of GdFeO$_3$ rotation
away from the ideal perovskite structure \cite{Ederer_07}. The hopping increases from 0.5eV to 0.65eV when the
GdFeO$_3$ rotation is decreased from the value observed for LaMnO$_3$ to zero (with the lattice parameters held constant).


To estimate the interaction strengths we follow the analysis of Kovaleva \emph{et al} \cite{Kovaleva_04} who
show that the optical spectrum of LaMnO$_3$ exhibits peaks which have a dependence on polarization and temperature allowing them 
to be associated to atomic-like excitations \cite{AhnMillis_00} with a reasonable degree of confidence. To establish notation 
and estimate uncertainties we give the details of the analysis here.
The formulation and results are similar to those given by Kovaleva \emph{et al} \cite{Kovaleva_04}, and the physical
arguments were introduced in Ref \cite{AhnMillis_00}.

We now turn to the local interactions.
We assume (as is apparently the case in the actual materials) that the crystal field (ligand field) 
is large enough that the $t_{2g}$ levels are well separated from the $e_g$ levels 
so that the pair hopping between $t_{2g}$ and $e_g$ orbitals is quenched (this assumption
was not made in \cite{Kovaleva_04} which accounts for the differences between their results and ours).
We also assume that the local interactions are strong enough that $t_{2g}$ level electrons are in their
maximum spin state, and may be treated as an electrically inert core spin of magnitude $|\vec{S}_c|=3/2$.
The on-site Hamiltonian in the $e_g$ manifold is then
\begin{eqnarray}
H_{loc} &=& \sum_{\sigma,\sigma'} (U- J) n_{1, \sigma} n_{2, \sigma'} + U \sum_{
i=1,2}n_{i,\uparrow} n_{i, \downarrow } + J( \, c^{\dagger}_{1,
\uparrow} c^{\dagger}_{1, \downarrow} c_{2, \downarrow} c_{2,
\uparrow} +h.c.)\nonumber \\ &-& 2 J \vec{s}_1 \cdot \vec{s}_2 
-2J_H \vec{S}_c \cdot (\vec{s}_1 + \vec{s}_2) + \Delta(n_1-n_2)
\label{eqn:H_loc}
\end{eqnarray}
Here $\vec{s}_i = \sum_{\alpha \beta} c^{\dagger}_{i \alpha} \vec{\sigma}_{\alpha \beta} c_{i \beta}$, $|\vec{S}_c|=3/2$
and $\Delta$ is a crystal field splitting of $e_g$ levels arising from a Jahn-Teller distortion of the Mn-O$_6$ octahedron
which may be static or dynamic, and have long ranged order or not. 
For a free ion $J_H=J$; we assume this henceforth because the general consensus is that
expect for $U$, intra atomic interactions are insensitive to solid state effect (screening). 
The eigenstates of $H_{loc}$ are characterized by the particle number $(n)$,
total spin ($S_{tot}$) and total $e_g$ spin ($S_{e_g}$), and the orbital configuration $(O)$. 
We label the two-electron states as $^{2 s_{e_g} + 1} O(S_{tot})$. There are 16 1-electron and 24 2-electron eigenstates,
taking the configurations of the core spin into account.

In the atomic picture, in the ground state of LaMnO$_3$ each Mn atom is in the state $n=1$, $S_{tot}=2$. The electron
is in the particular orbital state picked out by the crystal field splitting. An optical transition then leaves one site in the state 
$n=0$ $S_{tot}=3/2$ and one site in the state $n=2$ with orbital state labeled by $O$ and spin state 
characterized by $S_{tot}=5/2,3/2,1/2$ and $S_{e_g}=1,0$. Optical
peaks are at energies $\Delta E(S_{tot},S_{e_g},O) = E(n=2, S_{tot}, S_{e_g},O)+E(n=0,S_{tot}=3/2) - 2 E(n=1,S_{tot}=2, S_{e_g}=1/2) $.
Table I lists the 2-electron eigenstates, degeneracies and the corresponding optical transition energies.

\begin{center}
\begin{tabular}{|l|c|l|} \hline
States  & Degeneracy  & $\Delta E$    \\ \hline \hline
$^3A_2 (5/2)$ & 6 & $U-3J/2+2\Delta $       \\
$^3A_2 (3/2)$ & 4 & $U+7J/2+2\Delta $       \\
$^3A_2 (1/2)$ & 2 & $U+13J/2+2\Delta $      \\ \hline
$^1E^- (3/2)$ & 4 & $U+9J/2+2\Delta -\sqrt{4\Delta^2+J^2} $           \\
$^1A (3/2)$   & 4 & $U+9J/2+2\Delta +\sqrt{4\Delta^2+J^2} $            \\
$^1E^+ (3/2)$ & 4 & $U+7J/2+2\Delta $             \\ \hline
\end{tabular} \\ 
Table I: The 2-electron eigenstates of H (Eq(\ref{eqn:H_loc}))
la belled by $e_g$ spin (superscript), representation (letter with or without subscript) and
total spin degeneracy of the eigenstates (parenthesis),  and 
the corresponding optical  transition energies (see text)
\end{center}

The $^3A_2 (5/2)$ are the states of maximal spin and are favored by Hund's rule. They necessarily have one
electron in each orbital, hence a crystal field energy $2 \Delta$ higher than the starting state. The $^3A_2 (3/2,1/2)$
have the same orbital and $e_g$ spin configuration as $^3A_2 (5/2)$  but lower total spin. The $^3A_2 (1/2)$ state is not connected
to the ground state by the optical matrix element and will not be considered further here. 

The $^1E^+ (3/2)$ state is the low $e_g$ spin configuration with one electron in each orbital and the remaining 
two states $^1E^- (3/2)$ and $^1A (3/2)$ are $e_g$ singlets made up of linear combinations of states with
two electrons in the same orbital. These states are split by the crystal field but coupled by the inter-$e_g$ pair-hopping. 
If $2 \Delta >> J$, the pair hopping is quenched and we may identify the state $^1E^-$ as the coming from the two electron state
with both electrons in the orbital favored by the JT splitting and $^1A$ as the state with both electrons in the disfavored
orbital. The state $^1A$ is at high energy and is connected to the ground state by a very weak
matrix element; it will be disregarded. 
The relevant portion of excitation spectrum of the exact model therefore consists of 
transitions to the high spin, $^1E^-$  and $^1E^+$ states implying peaks at 
\be
\Delta E_{HS} &=& U-3J/2 + 2 \Delta \nonumber \\
\Delta E^{-}_{LS} &=& U+9J/2+ 2 \Delta  - \sqrt{4 \Delta^2 + J^2} \nonumber \\
\Delta E^{+}_{LS} &=& U+7J/2 + 2 \Delta \label{eqn:local_spectra}
\ee
Note that the latter two peaks are degenerate at $\Delta=0$.

An extensive theoretical literature exists on the problem of electrons coupled to classical core spins. We 
briefly discuss how to make comparison to these results. In the classical core-spin model we write the coupling
between core spin $\vec{S}_c$ and conduction spin $\vec{\sigma}_{el}$ ($|\vec{\sigma}_{el}|=1$) as 
$H_{cl} = -J_{cl} \vec{S}_c \cdot \vec{\sigma}_{el}$. The energy difference between high spin 
($\vec{\sigma}_{el}$ parallel to $\vec{S}_c$) and low spin ($\vec{\sigma}_{el}$ anti-parallel to $\vec{S}_c$)
is $2 J_{cl} S_c$. In the quantum model the energy differences depend on the total spin and electron number. The
high-spin/low-spin difference measured in optics is $5J$ while the difference in energy
between $n=1$ high-spin/low-spin is $4J$. Therefore there is an approximately 20$\%$ uncertainty in the classical
parameter $2 J_{cl} S_c$. 

The critical $U$ for a model with the interaction given in Eq(\ref{eqn:H_loc}) has not been calculated. However 
there are several limits in which the manganite model maps on to an effective one orbital model; for
these cases we may estimate the critical $U$. 
The first limit is of strong ferromagnetism. If the core spins are fully polarized the spin degree of freedom
is quenched, the orbital degree of freedom acts as a spin and the Jahn-Teller
coupling $\Delta$ as a magnetic field. Projection of Eq(\ref{eqn:H_loc}) on to the
maximum spin manifold then yields a one orbital Hubbard model with a $U_{eff} = U-3J/2$ 
(if $\Delta=0$). For this model the critical $U$ for the metal-insulator transition is $U_{c2}\sim1.5W\sim4.5$eV \cite{DMFT_96, Werner_07}.
A second simple limit is $J_H \rightarrow \infty$ (the $J_H$ in the actual materials is far from this limit).
In this case, in the paramagnetic phase the spin degree of freedom is again quenched and $U_{eff}$ is again $U-3J/2$.
The bandwidth is reduced by a factor of $\sqrt{2}$ \cite{AhnMillis_00,Michaelis_03} suggesting $U_{c2} \sim 1.5W /\sqrt{2} \sim 1.1W\sim3.3$eV. 


Experiment on orbitally ordered LaMnO$_3$ \cite{Kovaleva_04} identifies two clear peaks 
at 2eV and 4.4eV, along with weaker features at 4.7eV and higher energies. 
In this material it is reasonable to regard the Jahn-Teller distortion $\Delta$ as frozen-in.
For $T<T_N \sim 140K$ LaMnO$_3$ is an A-type
antiferromagnet with ferromagnetic planes, which we take to define the $x-y$ plane, antiferromagnetically
alternating in the remaining, $z$, direction. Comparison of spectra taken with electric field along and perpendicular
to $z$ and at temperatures above and below $T_N$ implies \cite{Kovaleva_04,AhnMillis_00} that the 2eV peak corresponds to the HS (high-spin)
final state and the 4.4eV to a LS (low-spin) final state. There is only one HS state so we identify
$U-3J/2 + 2\Delta \sim 2$eV. There are two candidate LS states and therefore there is an uncertainty in the
peak assignment. Consideration of the optical transition strengths implied by the observed orbital order
suggests that the $E^{+}_{LS}$ state should be more prominent implying
we identify 2.4eV=$\Delta E^{+}_{LS} - \Delta E_{HS}=5J$ so that $J=0.48$eV and $U+2\Delta \sim 2.7$eV. 
In this interpretation the transition to the $\Delta E^{-}_{LS}$ state would give rise to a weaker 
feature below the mean peak, not resolved as a separate excitation
because of the band-broadening. The observed approximate 0.5eV width then implies that the $\Delta E^{JT}_{LS}$ cannot be more than about
0.5eV below $\Delta E^{\bar{JT}}_{LS}$ implying $2\Delta \lesssim 0.7$eV and 2.0eV$<U<$2.7eV. 
Alternatively we may identify the 4.4eV with $E^{-}_{LS}$ implying 2.4eV$\sim 6J-\sqrt{4 \Delta^2 + J^2}$.
We would further identify the 4.7eV peak with $E^{+}_{LS}$ 
implying $J\sim 0.54$eV, $2\Delta \sim 0.64$eV, and $U \sim 2.18$eV. The ambiguity in peak assignment
therefore does not affect our estimates of the interaction parameters.

These values are reasonably consistent with the gas phase Mn value $J\sim 0.5$eV and with the band calculation \cite{Ederer_07}
which suggests $2 \Delta \sim 0.53$eV and $J\sim 0.65$eV (note that in \cite{Ederer_07} the energy difference
for 1 electron with spin up or spin down is $2J$ with the fit $J \sim 1.3$eV, 
while from Eq(\ref{eqn:H_loc}) the difference is $4J$ due to $|\vec{S}_c| = 3/2$, therefore one has to divide the $J$ in \cite{Ederer_07}
by 2 to compare the $J$ fitted here).
To summarize, the data and other information are consistent with the estimates $U=2.3 \pm 0.3$eV, $2\Delta \sim J \sim 0.5$eV.

\begin{figure}[htbp]
   \centering
   \epsfig{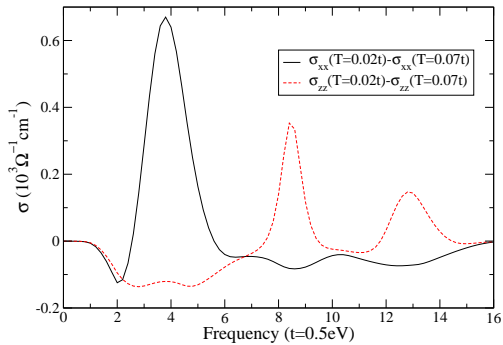}
   \caption{(Color online) The change in optical conductivity across the Neel transition. The states at temperature $T=0.02t$ and  
	$T=0.07t$ are A-type antiferromagnetic and paramagnetic respectively \cite{cLin_08}. The plot is designed to be
	directly compared to the measurement reported in Ref\cite{Kovaleva_04}, Fig(2).
	}
   \label{fig:dsigma}
\end{figure}

The analysis of the optics given above was based on the atomic limit. 
We have employed the semiclassical solver devised by Okamoto $et$ $al$ \cite{Okamoto_05} to
solve the model specified by Eq(\ref{eqn:H_loc}), using the tight binding dispersion from Ref\cite{Ederer_07,cLin_08}
with the bandwidth $W=6t=3$eV implied by the band theory calculations of Ref\cite{Ederer_07}. In Fig(\ref{fig:dsigma})
we show our calculated results \cite{cLin_08} for the change in conductivity across the Neel transition.
The results are presented in a form which allows direct comparison
to the experimental results of Kovaleva et al \cite{Kovaleva_04}. The excellent agreement of
energy scales and reasonable agreement of form and magnitude confirm the validity of the atomic limit analysis.

\begin{figure}[ttt]
   \centering
   \epsfig{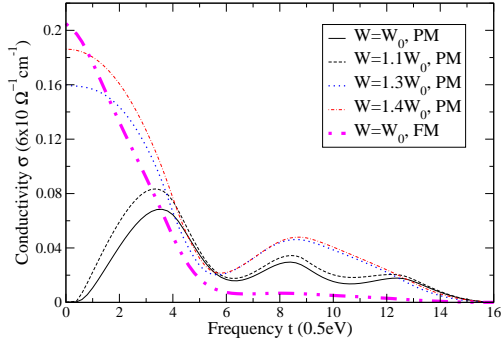}
   \caption{ (Color online) The optical conductivities for bandwidth ranging from $W=W_0=3$eV (solid) to $W=1.4W_0$ 
	(dash-double dot). All curves
	are computed at temperature $T=0.04t$ which is roughly 30$\%$ the calculated orbital ordering temperature,
	i.e. roughly of order room temperature. 
	The heavy dash-double dotted curve is calculated for a hypothetically ferromagnetic, orbitally disordered phase,
	all others are calculated for the paramagnetic but orbitally disordered phase.
	}
   \label{fig:DOS}
\end{figure}

We have used the semiclassical approximation to calculate properties for different temperatures, bandwidths and
interaction strengths. Representative results are shown in Fig(\ref{fig:DOS}). 
Reducing the GdFeO$_3$ distortion is equivalently increasing the bandwidth $W$.
Forcing the solution to be orbitally disordered, we found that in the paramagnetic phase the system displays a gap at low temperature $T=0.04t$ for $W=W_0$ (solid) and $1.1W_0$ (dashed) indicating a Mott insulating state. When the bandwidth is $W=1.3 W_0$ (dotted) or larger the
system becomes metallic. We also 
consider a ferromagnetic state which has an effective bandwidth roughly 1.4 ($\sqrt{2}$) times larger than that of the paramagnetic phase
\cite{AhnMillis_00,Michaelis_03}.
We see that for bandwidth $W=W_0$, the system is metallic for ferromagnetic, orbitally disordered phase
(heavy dash-double dotted curve in Fig(\ref{fig:DOS})) which is consistent with the results from directly varying the bandwidth
to $W=1.4W_0$ (dash-double dotted curve) in the paramagnetic phase. 
 Our calculation thus indicates that LaMnO$_3$ $is$ a Mott insulator with the
local interaction strength very close to but slightly stronger than the critical value of Mott transition,
and also indicates that ferromagnetic and the large $J_H$ limits discussed above provide a poor representation of the physics.
Finally we note that if the electron-lattice is ``turned off'', i.e. setting $\Delta=0$ in Eq(1), 
the material becomes metallic (curve not shown).

At this point we it is useful to discuss more carefully what is meant by the term ``Mott insulator''
in the multi-orbital LaMnO$_3$ context. The $e_g$ manifold has a four-fold local degeneracy
(2x spin and 2x orbital) so an ordering with at least a 4-site unit cell would be required
to produce a ``Slater'' insulator. A priori one could discuss
 a Mott transition in the fully symmetric (4x local degeneracy) case, or in the partially ordered case
(2x local degeneracy) where only one of the spin and orbital symmetries is broken. However 
our calculations indicated that in the manganites
the spin states are strongly split so the low
energy physics is characterized only by a 2-fold orbital degeneracy (along with
the global configuration of core spins). The criterion for Mott insulator
is that the effective inter-orbital interaction within the high-spin, orbitally
degenerate manifold is large enough to open a gap at zero temperature within the single site DMFT approximation.
We find that in the paramagnetic, orbitally disordered phase the interaction is slightly larger
than the critical value. However this interaction is composed of two physically distinct contributions,
one from interorbital Coulomb interaction and the other from the electron-lattice coupling. Removing
the contribution from the electron-lattice coupling changes the interaction strength from
slightly larger than the critical value to slight smaller. While in the literature
the term ``Mott insulator'' is used for Coulomb-driven phenomena, we think
it is appropriate to use it for insulating behavior driven by any local interaction, and therefore we identify
LaMnO$_3$ as being a Mott insulator.

We emphasize that our findings place LaMnO$_3$ close to the edge of the insulating phase boundary. For this reason details
including the presence and evolution with doping of GdFeO$_3$ rotation (which changes the bandwidth
at the $30\%$ level) as well as the possible presence of ferromagnetic order (leading to a 40$\%$ increase in bandwidth)
become very important. Finally we note that the semiclassical method used here places the transition
of the one orbital Hubbard model at $U=U_{c1}$. In the one orbital Hubbard model the difference between
$U_{c1}$ and $T=0$ metal-insulator point $U_{c2}\sim 1.2 U_{c1}$ is due to the formation of a Kondo
resonance, which is expected to be suppressed by the core spin coupling in the present model. However, further investigation
of this point would be desirable.

\section*{Acknowledgment}
We acknowledge support from DOE-ER 46169 and Columbia MRSEC.



\end{document}